\documentclass[prd,
twocolumn,
superscriptaddress,nofootinbib,%
tightenlines,showpacs,showkeys]{revtex4}

\usepackage{amsfonts,amsmath,graphicx,epsfig,amssymb}
\usepackage{epsfig}
\usepackage{mathrsfs}
\usepackage{latexsym}
\usepackage{amsxtra}
\usepackage{amsbsy}
\usepackage{amscd}
\usepackage{color}
\usepackage{bbm}
\usepackage{hyperref}
\usepackage{multirow}

\usepackage{graphics}
\usepackage{amsfonts}

\usepackage{subfigure}
\usepackage{color}
\allowdisplaybreaks

\newcommand{\rbox}{\rule[-0.20cm]{0cm}{5mm}}

\newcommand{\be}{\begin{equation}}
\newcommand{\ee}{\end{equation}}

\def\no{\nonumber}
\def\bea{\arraycolsep .1em \begin{eqnarray}}
\def\eea{\end{eqnarray}}

\begin{document}
\title{\Large A study on the correlation between poles and cuts in $\pi\pi$ scattering}
\author{Ling-Yun Dai}
\email{dailingyun@hnu.edu.cn}
\affiliation{School of Physics and Electronics, Hunan University, Changsha 410082, China}
\author{Xian-Wei Kang}
\email{kangxianwei1@gmail.com}
\affiliation{College of Nuclear Science and Technology, Beijing Normal University, Beijing 100875, China}
\author{Tao Luo }
\email{luot@fudan.edu.cn}
\affiliation{Key Laboratory of Nuclear Physics and Ion-beam Application
(MOE) and Institute of Modern Physics, Fudan University, Shanghai 200443, China}
\author{Ulf-G. Mei{\ss}ner }
\email{meissner@hiskp.uni-bonn.de}
\affiliation{Helmholtz Institut f\"ur Strahlen- und Kernphysik and Bethe Center
 for Theoretical Physics, Universit\"at Bonn, D-53115 Bonn, Germany}
\affiliation{Institute for Advanced Simulation, Institut f\"ur Kernphysik
   and J\"ulich Center for Hadron Physics, Forschungszentrum J\"ulich, D-52425 J\"ulich, Germany}
\begin{abstract}
In this paper we propose a dispersive method to describe two-body
scattering with unitarity imposed. This approach is applied to elastic $\pi\pi$
scattering. The amplitudes keep single-channel unitarity and
describe the experimental data well, and the low-energy amplitudes
are consistent with that of chiral perturbation theory. The pole
locations of the $\sigma$, $f_0(980)$, $\rho(770)$ and $f_2(1270)$ and
their couplings to $\pi\pi$ are obtained. A virtual state appearing
in the isospin-two S-wave is confirmed. The correlations between the left
(and right) hand cut and the poles are discussed. Our results show
that the poles are more sensitive to the right hand cut rather than
the left hand cut.
The proposed method could be used to study other two-body scattering processes.
\end{abstract}

\maketitle

\parskip=2mm
\baselineskip=3.5mm

\section{Introduction}\label{sec:introduction}
In a two-body scattering system, for example two hadrons, the general
principals that we know are unitarity, analyticity, crossing,
the discrete symmetries, etc. The resonances that appear as the
intermediate states in such system are important. Among them the
lightest scalar mesons, related to  $\pi\pi$ scattering, have the same
quantum numbers as the QCD vacuum and are rather
interesting, for some early references, see~\cite{MRP10,jaffe4q,Meissner:1990kz}.
The $\pi\pi$ scattering amplitude is also
crucial to clarify the hadronic contribution to the anomalous
magnetic moment of the muon, see e.g.~\cite{Colangelo:2018mtw,Danilkin:2018qfn}.
To study the resonances in a given scattering process, one needs dispersion
relations to continue the amplitude
from the real $s$-axis  (the physical region) to the complex-$s$
plane~\cite{Kang:2013jaa,DLY-MRP14,Chen2015,Hanhart:2016pcd}, where
the pole locations and their couplings are extracted.
Following this method, some work on the light scalars can be found
in~\cite{zheng00,colangelo01,zhou04,Moussallam06,caprini06,PelaezPRL},
where the accurate pole locations and residues of the $\sigma$
and $\kappa$ mesons are given.

For the dispersive methods, a key problem is how to determine the left hand cut
(l.h.c.) and the right hand cut (r.h.c.), with the unitarity kept at the same time.
In Refs.~\cite{DLY-MRP14,Chen2015} the l.h.c is estimated by crossed-channel exchange of resonances,
where chiral effective field theory ($\chi$EFT)
is used to calculate the amplitude. And the contribution of r.h.c. is represented by an
Omn\`es function, with unitarity kept.
In the well-known Roy equations, crossing symmetry and analyticity are perfectly combined together
as the l.h.c is represented by the unitary cuts of the partial waves. The single channel unitarity is also
well imposed by keeping the real part of the partial wave amplitudes the same as what is calculated
by the phase shift directly, which could be obtained by fitting to the experimental data in some analyses.
Until now, Roy and Roy-Steiner equations certainly give the most accurate description of
the two-body scattering
amplitude and the information of resonances appearing as the intermediate states, such as
$\pi\pi$, $\pi K$ scattering and the pole locations and residues of the $\rho$, $\sigma$,
$f_0(980)$ and $\kappa$, etc., see e.g.~\cite{Moussallam06,caprini06,PelaezPRL}.
In addition, Ref.~\cite{caprini06} shows that the l.h.c. can not be ignored for the determination of
the pole location of the $\sigma$.
By removing the parabola term of the l.h.c., the $\sigma$ pole location is changed by about
15\% accordingly,
while the unitarity is violated due to the removal of the l.h.c.. And thus the method to get
the poles on the second Riemann sheet, calculated from the zeros of the S-matrix, is not
reliable any more, as the method is based on the continuation implemented by unitarity.
Here, we focus on obtaining a quantitative relation between cuts and poles, with unitarity imposed and
the l.h.c. and r.h.c. are correlated with each other.

This paper is organized as follows: In Sect.~II we establish a
dispersive method based on the phase. In the physical region we also
represent the amplitudes by an Omn\`es function of the phase above threshold.
In Sect.~III we fit the $\pi\pi$ scattering amplitudes up
to 1~GeV in a model-independent way, including the $IJ=00,02,11,20$
waves, where $I$ denotes the total isospin and $J$ the angular momentum.
The fit results are the same as those given by the Omn\`es
function representation and comparable with those of chiral
perturbation theory ($\chi$PT) in the low-energy region. The poles
and couplings are also extracted. In Sect.~IV we give the estimation
of the relation between poles and cuts, including both the l.h.c and
the r.h.c.\ . We end with a brief summary.

\section{Scattering amplitude formalism}\label{sec:formalism}
\subsection{A dispersive representation}\label{sec:sub;single}
The two-body scattering amplitude can be written as:
\bea
T(s)= f(s) e^{i \varphi(s)}\,, \label{eq:T}
\eea
with $\varphi(s)$ the phase and $f(s)$ a real function. By writing a dispersion relation for
$\ln T(s)$, one has:
\bea
\ln T(s)&=&\ln f(s_0) +\frac{s-s_0}{\pi}\int_L \frac{\varphi_L(s') ds' }{(s'-s_0)(s'-s)}\no\\
&+&\frac{(s-s_0)}{\pi}\int_R \frac{\varphi_R(s') ds'
}{(s'-s_0)(s'-s)}\,.\label{eq:T;DR}
\eea
Here, $s_0$ is chosen at a specific point
where the amplitude is real, and \lq L' denotes the l.h.c. and \lq
R' stands for the r.h.c.. The amplitude turns into
\bea
T(s)=T(s_0)\Omega_L(s)\Omega_R(s)\,.\label{eq:T;Omnes}
\eea
On the other hand, unitarity is a general principal required for the scattering
amplitude. In the single channel case one has
\bea {\rm Im}
T(s)=\rho(s) |T(s)|^2\, ,\label{eq:unitarity}
\eea
where $s$ is in the elastic region and $\rho(s)$ is the phase space factor. Substituting
Eq.~(\ref{eq:T;Omnes}) into Eq.~(\ref{eq:unitarity}),
we obtain a representation (in the elastic region) for a single
channel scattering amplitude
\bea
T(s)=-\frac{{\rm Im}[\Omega_R(s)^{-1}]\Omega_R(s)}{\rho(s)}\,.\label{eq:unitarity;ph}
\eea
Also, the Omn\`es function of the phase for the l.h.c. is correlated
with that of the r.h.c.
\bea \Omega_L(s)=-\frac{{\rm
Im}[\Omega_R(s)^{-1}]}{\rho(s) T(s_0)}\,,\label{eq:unitarity;lhc}
\eea
which is again valid in the elastic region.
A simple way to get the two-body scattering amplitude proceeds in two steps:
First, we follow Eq.~(\ref{eq:unitarity;ph})
to fit the Omn\`es function of the r.h.c. to  experimental data, and
then use Eq.~(\ref{eq:unitarity;lhc}) and
other constraints below the threshold to fit the Omn\`es function of the l.h.c.\ .
Note that Eq.~(\ref{eq:unitarity;ph}) does not only work for the single channel
case, but also for the coupled channel case in the physical region.

\subsection{On $\pi\pi$ scattering}\label{sec:sub;th}
In the equations above, the threshold factor is not included. Considering such factors, we need
to change the amplitudes into:
\bea
T^I_J(s)=(s-z^I_J)^{n_J}f^I_J(s)e^{i\varphi^{IJ}(s)}\,.\label{eq:T;th}
\eea
Here and in what follows, we take $\pi\pi$ scatering as an example.
Thus one has $z^I_J=4 M_\pi^2$ for the P-, D-, and higher partial waves, and
$z^I_J$ is the Adler zero for the S-waves.
$n_J$ is one for S- and P-waves and two for D waves.
We define a reduced amplitude
\bea
\tilde{T}^I_J(s)=f^I_J(s)e^{i\varphi^{IJ}(s)}\,,\label{eq:T;red}
\eea
and again we can write a dispersion relation for $\ln\tilde{T}^I_J(s)$, so that we have
\bea
\ln \tilde{T}^I_J(s)&=&\ln f^I_J(s_0) +\frac{s-s_0}{\pi}\int_{-\infty}^{0} \frac{\varphi^{IJ}_L(s') ds' }{(s'-s_0)(s'-s)}\no
\\&+&\frac{(s-s_0)}{\pi}\int_{4M_\pi^2}^{\infty} \frac{\varphi^{IJ}_R(s') ds' }{(s'-s_0)(s'-s)}\,.\label{eq:T;DR;red}
\eea
Here, $s_0$ could be chosen from the range $[0,4 M_\pi^2]$. For the r.h.c., we cut off the integration
somewhere in the high energy region, see discussions in the next sections.
We have
\bea
f^I_J(s_0)=\frac{T^I_J(s_0)}{(s_0-z^I_J)^{n_J}}\,, \no
\eea
and
\bea
T^I_J(s)=T^I_J(s_0)\left(\frac{s-z^I_J}{s_0-z^I_J}\right)^{n_J}\Omega^{IJ}_L(s)\Omega^{IJ}_R(s)\,.\label{eq:T;Omnes;red}
\eea
The $T^I_J(s_0)$ could be fixed by $\chi$PT or scattering lengths, or other low-energy constraints.
For simplicity, we choose $s_0=0$.
Combining  unitarity, embodied by Eq.~(\ref{eq:unitarity;ph}),  we have a correlation
between Omn\`es functions of l.h.c. and r.h.c. in the elastic region
\bea
\Omega^{IJ}_L(s)=-\frac{{\rm Im}[\Omega^{IJ}_R(s)^{-1}](s_0-z^I_J)^{n_J}}{\rho(s) T^I_J(s_0)(s-z^I_J)^{n_J}}\,.\label{eq:unit;lhc}
\eea
This is similar to Eq.~(\ref{eq:unitarity;lhc}). Substituting Eq.~(\ref{eq:unit;lhc}) into
Eq.~(\ref{eq:T;Omnes;red}), we still have  Eq.~(\ref{eq:unitarity;ph}).
Since we know the $\pi\pi$ scattering amplitudes well in the region [$4M_\pi^2$, 2~GeV$^2$] and $\chi$PT
describes the amplitudes well in the low-energy region, we have to fit the l.h.c. to
both Eq.~(\ref{eq:unitarity;lhc}) and $\chi$PT.

\section{Phenomenology }
\subsection{Fits}
For the $\pi\pi$ scattering amplitude, we can parametrize the phase caused by the
l.h.c. by a conformal mapping
\bea
\varphi^{IJ}_L(s)=\sum_{n=1}^k c^{IJ}_n {\rm Im}[\omega(s)]^n \,,\label{eq:phiL}
\eea
with
\bea
\omega(s)=\frac{\sqrt{s^3}-\sqrt{(s^{IJ}_L)^3}}{\sqrt{s^3}+\sqrt{(s^{IJ}_L)^3}}\,. \label{eq:comformal}
\eea
Notice that  ${\rm Im}~\omega(s)$ behaves as $\sqrt{-s^3}$ around $s=0$, which is consistent with
that of$\chi$PT, see Ref.~\cite{zhou04} and references therein.

As concerns  the r.h.c., it is less known in the high energy region. However,
these distant r.h.c. have less important effects in the low-energy
region, especially in the region $s\leq 1$~GeV$^2$. We choose three
kinds of $\Omega^{IJ}_R(s)$ to test the stability and uncertainty
caused by the distant r.h.c.. In Case~A, the phases~\cite{DLY-MRP14} are
cut off at $s=2.25$~GeV$^2$. In Case~B, the phases are given
by~\cite{DLY-MRP14}, up to $s=22$~GeV$^2$. In Case~C, the
phases/Omn\`es functions of the r.h.c. are given
by~\cite{Dai:2017tew,Dai:2017uao} and references
therein, up to $s=22$~GeV$^2$. Here, the phases are fitted to
the experimental data~\cite{CERN-Munich,OPE1973} up to
$\sqrt{s}=2$~GeV and constrained by unitarity up to
$\sqrt{s}=4$~GeV. Notice that in Case~A and B the phase of the isospin-one
P-wave is given by CFDIV~\cite{KPY}, and we continue it to the
higher energy region by means of the  function
\bea
\varphi^{11}_{R,h}(s)=\varphi^{11}_{\infty}+B[k,n]\left(\frac{s_R}{s}\right)^k+C[k,n]
\left(\frac{s_R}{s}\right)^n\,,
\eea
with
\bea
B[k,n]&=&\varphi^{11}_{R}(s_R)-\varphi^{11}_{\infty}-C[k,n]\,,\no\\
C[k,n]&=&\frac{k[\varphi^{11}_{\infty}-\varphi^{11}_{R}(s_R)]-{\varphi'}^{11}_{R}(s_R)s_R}{n-k}\,.
\eea
The function (and also its first derivative) is smooth at the
point $s_R$. We set $k=1$, $n=2$, $s_R=1.4^2$~GeV$^2$ and
$\varphi^{11}_{\infty}=160^\circ$, which is close to
$\varphi^{11}_R((1.4\, \text{GeV})^2)=170^\circ$ and ensures that the
phase in the high energy region behaves smoothly. The upper limits of
the integration of the r.h.c. of isospin-one P-wave are the same as
the other partial waves.

\begin{figure}[hpt]
\includegraphics[width=0.48\textwidth,height=0.35\textheight]{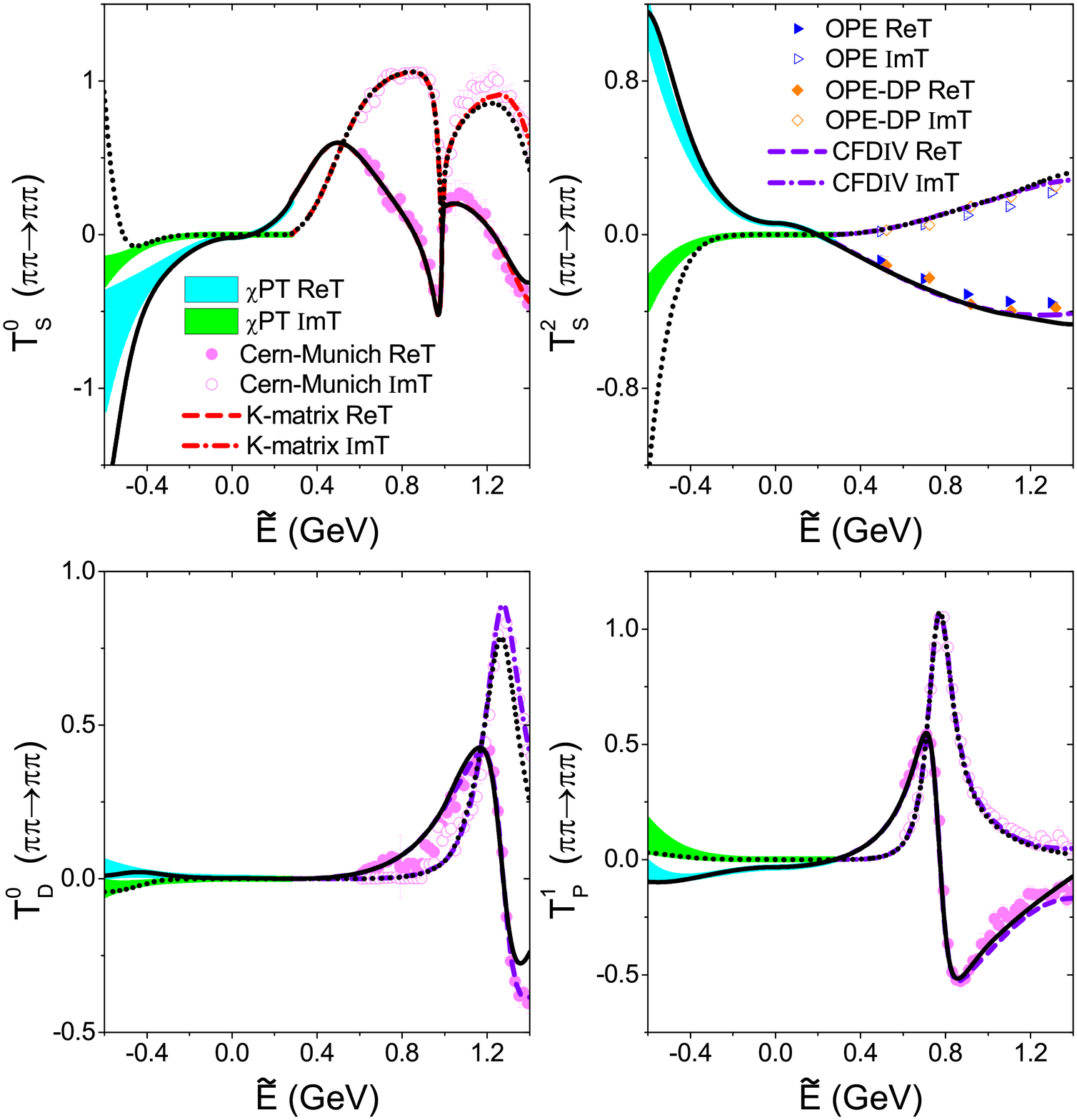}
\caption{\label{Fig:T} Fit of the $\pi\pi$ scattering amplitudes for Case B. Notice that
  $\tilde{E}=\rm{sgn}(s)\sqrt{s}$. The solid lines denote the real part of the amplitudes and the dotted, dashed, dash-dotted and dash-dot-dotted lines denote the imaginary part.
  The black lines are from our fit. The red lines are from a K-matrix fit \cite{DLY-MRP14} for the isospin-zero S-wave, and the violet
  lines are from CFDIV~\cite{KPY} for other waves.
  The borders of the cyan and green bands in the low-energy region are from
  $SU(2)$ and $SU(3)$ $\chi$PT, respectively.  The CERN-Munich data are from Ref.~\cite{CERN-Munich},
  and the OPE and OPE-DP data are from~\cite{OPE1973}.}
\end{figure}

The parameters of our fits for all the Cases are given in Tab.~\ref{tab:para}.
\begin{table}[htbp]
\hspace{-1.5cm}
\vspace{-0.0cm}
{\footnotesize
\begin{center}
\tabcolsep=0.11cm
\begin{tabular}  {|c|c|c|c|c|c|}
\hline
\multirow{2}{*}{\rule[-0.8cm]{0cm}{1.6cm}} & \multirow{2}{*}{\rule[-0.5cm]{0cm}{1.0cm}Case A}      &  \multirow{2}{*}{\rule[-0.5cm]{0cm}{1.0cm}Case B}     & \multirow{2}{*}{\rule[-0.5cm]{0cm}{1.0cm}Case C}    & \multicolumn{2}{c|}{$\chi$PT}\\
\cline{5-6}
  &       &       &      & $SU(2)$  & $SU(3)$ \\[0.5mm]
\hline
$s^{00}_L$      & 1.0        &  0.5        & 0.85       & -  & -  \\
$c^{00}_1$      & 4.8161     &  1.4158     & 2.4397     & -  & -  \\
$c^{00}_2$      & 3.6241     &  0.9454     & 1.5154     & -  & -  \\
\rule[-0.25cm]{0cm}{0.5cm}$T^0_S(0)$       &-0.02080    &-0.02080    & -0.02080   & -0.016  & -0.016      \\ \hline
$s^{02}_L$      & 2.0        &  0.4        & 2.5      & -  & -    \\
$c^{02}_1$      & 5.5424     &  1.3611     & 6.2776   & -  & -    \\
\rule[-0.25cm]{0cm}{0.5cm}$T^0_D(0)$ & 0.0030  & 0.0020  & 0.0030    & 0.0035  & 0.0032     \\ \hline
$s^{11}_L$      & 1.0        &  1.0        & 0.85     & -  & -      \\
$c^{11}_1$      & 0.5681     & -1.0873     & 0.4523    & -  & -   \\
$c^{11}_2$      & -          & -0.9923     &-0.3598    & -  & -   \\
\rule[-0.25cm]{0cm}{0.5cm}$T^1_P(0)$       &-0.03297   & -0.03297  &-0.03550   & -0.034  & -0.034   \\ \hline
$s^{20}_L$      & 1.3        &  1.3        & 1.6       & -  & -   \\
$c^{20}_1$      & 4.3718     &  5.6436     & 3.9488   & -  & -    \\
$c^{20}_2$      & 1.5917     &  2.6051     & -        & -  & -   \\
\rule[-0.25cm]{0cm}{0.5cm}$T^2_S(0)$       & 0.055       &  0.055      & 0.060    & 0.058  & 0.057      \\ \hline
\hline
\end{tabular}
\caption{\label{tab:para} The parameters for each fit. ``-'' means
the absence of the corresponding quantity. For comparison, we also give the one-loop $\chi$PT results.}
\end{center}
}
\end{table}
The $c^{IJ}_n$ are determined by the following procedure. In the elastic region, we choose one or
two \lq mesh points', depending on how many coefficients $c^{IJ}_n$ we require.
Combining Eqs.~(\ref{eq:unit;lhc},\ref{eq:phiL}), we can build a matrix and solve for $c^{IJ}_n$.
This strategy gives a good description of the amplitudes, with unitarity kept.
See the fit results shown in Fig.~\ref{Fig:T}.
Here all the partial waves refer to  Case~B, in which the phase is cut off at $s=2.25$~GeV$^2$.
The amplitudes from the other Cases are quite close to this one, except for the inelastic region and
the distant l.h.c. ($\tilde{E}\leq-0.4$~GeV$^2$).
Our fit, both the real part (black solid line) and imaginary (black dotted line) part of the
amplitudes shown in Fig.~\ref{Fig:T}, is
indistinguishable from that given by the K-Matrix \cite{DLY-MRP14} or CFDIV \cite{KPY}.
Note that the amplitudes given by Eq.~(\ref{eq:unitarity;ph}) are exactly the same as
those of the K-Matrix or CFDIV
from $\pi\pi$ threshold to the inelastic threshold. This implies that the unitarity is
respected. To test it quantitatively, we define
\bea
\mathcal{R}_{ T^I_J}&=&\frac{1}{N}\sum_{n=1}^N\frac{|\Delta T^I_J(s_n)|}{|T^I_J(s_n)|}\;.
\label{eq:Delta;T}
\eea
$\Delta T^I_J(s_n)$ is the difference between our amplitude and that of Eq.~(\ref{eq:unitarity;ph}).
Here we choose $s_n=0.1-0.9$~GeV$^2$ for the S-waves, $s_n=0.1-0.8$~GeV$^2$ for the P-wave, and
$s_n=0.1-1.0$~GeV$^2$ for the D-wave, with step of 0.1~GeV$^2$. These points are located
between the $\pi\pi$ and the inelastic thresholds.
From here on all the steps are chosen to be 0.1~GeV$^2$ (or 0.1~GeV for $\tilde{E}$).
We find that $\mathcal{R}_{ T^0_S}=0.1\%$, $\mathcal{R}_{ T^1_P}=0.1\%$, $\mathcal{R}_{ T^2_S}=1.4\%$,
and $\mathcal{R}_{ T^0_D}=1.4\%$. The violation of unitarity is rather small.

For $T^I_J(0)$, $\chi$PT could be used to fix it. The analytical
$SU(3)$ 1-loop $\chi$PT amplitudes of each partial waves, are
recalculated and given in Appendix.~\ref{app:ChPT}. The low-energy
constants are given by \cite{Bijnens:2014lea}.
Those of $SU(2)$ 2-loop $\chi$PT amplitudes are given by
\cite{Bijnens:1997vq,zhou04} and references therein. All the values
of $T^I_J(0)$ in Tab.~\ref{tab:para} are close to the prediction of
$\chi$PT or our earlier analyses \cite{Dai:2017tew,Dai:2017uao}.
In the isospin-zero S-wave, the magnitude of our $T^0_0(0)$ is a bit larger than that of $\chi$PT.
This is consistent with what is known about this  scattering length, where the one-loop
$\chi$PT calculation gives a smaller result than what is obtained by dispersive methods, Roy equations or in experiment,
see e.g. the review~\cite{Bijnens:2014lea}. A better comparison would be given with the 2-loop $\chi$PT amplitudes.

In the isospin-zero
D-wave, the $T^0_2(0)$ varies more in the different Cases. The reason is that some fine-tuning
is needed as the inelastic r.h.c. is difficult to be implemented well. The amplitudes
given by Eq.~(\ref{eq:unitarity;ph}) are much different from that of
CFDIV in the inelastic region where the $f_2(1270)$ appears. Notice further
that the value of $T^0_D(0)$ is very small, one order smaller than that of the other waves.

\subsection{Pole locations and couplings}
With these amplitudes given by a dispersion relation, the information of the poles
can be extracted. The pole $s_R$ and its coupling/residue $g_{f\pi\pi}$ on the second Riemann
sheet are defined as
\bea
T^{II}(s)=\frac{g_{f\pi\pi}^{2}}{s_R-s}\,.\label{eq;g}
\eea
Note that the continuation of the $T(s)$ amplitude to the second Riemann sheet is based on
unitarity,
\bea
T^{II}(s+i\epsilon)=T^{I}(s-i\epsilon)=\frac{T^{I}(s+i\epsilon)}{S^{I}(s+i\epsilon)}\,.\label{eq;T;II}
\eea
The poles and couplings/residues for  Cases~A,B,C are given in Table \ref{tab:poles;case}.
\begin{table}[hpt]
{\footnotesize
\begin{tabular}{|@{}c @{}|c | c |c |@{}c @{}|@{}c@{} |}
\hline
\rule[-0.4cm]{0cm}{0.8cm}\multirow{2}{*}{\rule[-0.8cm]{0cm}{1.6cm}State} & \rule[-0.4cm]{0cm}{0.8cm}\multirow{2}{*}{\rule[-0.8cm]{0cm}{1.6cm}Case} &  pole locations  &\multicolumn{2}{c|}{$g_{f\pi\pi}=|g_{f\pi\pi}|e^{i\phi}$~}  \\
\cline{3-5}
\rule[-0.4cm]{0cm}{0.8cm}   &  & (MeV)   & $~|g_{f\pi\pi}|~(GeV)~$ & $~\phi~(^\circ)~$   \\
\hline\hline
\multirow{3}{*}{\rule[-1cm]{0cm}{2cm}$\sigma/f_0(500)$} & A & $432.5-i269.8$
     & $0.46$   & $-77$      \rbox \\[0.5mm] \cline{2-5}
     \multirow{3}{*}{\rule[-1cm]{0cm}{2cm}}  & B & $442.7-i270.5$
     & $0.48$   & $-74$      \rbox \\[0.5mm]  \cline{2-5}
     \multirow{3}{*}{\rule[-1cm]{0cm}{2cm}} & C & $438.2-i270.6$
     & $0.47$   & $-75$      \rbox \\[0.5mm] \hline
\multirow{3}{*}{\rule[-1cm]{0cm}{2cm}$f_0(980)$}   & A       & $997.5-i19.0$
     & $0.25$   & $-81$    \rbox \\[0.5mm] \cline{2-5}
\multirow{3}{*}{\rule[-1cm]{0cm}{2cm}}      & B   & $997.6-i21.6$
     & $0.27$   & $-83$    \rbox \\[0.5mm] \cline{2-5}
\multirow{3}{*}{\rule[-1cm]{0cm}{2cm}}      & C   & $997.6-i20.5$
     & $0.26$   & $-82$    \rbox \\[0.5mm] \hline
\multirow{3}{*}{\rule[-1cm]{0cm}{2cm}$f_2(1270)$ }   & A    & $1260.9 -i111.2$
        & $0.55$   & $-10$    \rbox \\[0.5mm] \cline{2-5}
\multirow{3}{*}{\rule[-1cm]{0cm}{2cm} }   & B    & $1294.1 -i57.9$
        & $0.52$   & $11$    \rbox \\[0.5mm] \cline{2-5}
\multirow{3}{*}{\rule[-1cm]{0cm}{2cm} }   & C    & $1266.0 -i99.5$
        & $0.54$   & $-8$    \rbox \\[0.5mm] \hline
\multirow{3}{*}{\rule[-1cm]{0cm}{2cm}$\rho(770)$  }  & A      & $761.1 -i70.6 $
    & $0.34$   & $-12$                     \rbox \\[0.5mm] \cline{2-5}
\multirow{3}{*}{\rule[-1cm]{0cm}{2cm}}    & B    & $763.0 -i73.3 $
    & $0.35$   & $-11$                     \rbox \\[0.5mm] \cline{2-5}
\multirow{3}{*}{\rule[-1cm]{0cm}{2cm}}     & C   & $761.3 -i71.7 $
    & $0.34$   & $-12$                     \rbox \\[0.5mm] \hline
\multirow{3}{*}{\rule[-1cm]{0cm}{2cm}$2S~~v.s.$}   & A     & $29.8  $
    & $9.8\times10^{-3}$   & $90$                      \rbox \\[0.5mm] \cline{2-5}
\multirow{3}{*}{\rule[-1cm]{0cm}{2cm}}    & B    & $29.8  $
    & $9.8\times10^{-3}$   & $90$                      \rbox \\[0.5mm] \cline{2-5}
\multirow{3}{*}{\rule[-1cm]{0cm}{2cm}}    & C    & $32.3  $
    & $11.0\times10^{-3}$   & $90$                      \rbox \\[0.5mm] \hline
\end{tabular}
\caption{\label{tab:poles;case}The pole locations and residues given
by our fits. The notation ``$2S~~v.s.$'' denotes the virtual state in
the isospin-two $S$-wave. }
}
\end{table}

All the poles and residues of the different Cases are close to each
other, except for the pole location of the $f_2(1270)$. The reason
is that the $f_2(1270)$ is located outside the elastic unitary cut of
$T^0_D(s)$, while Eq.~(\ref{eq:unit;lhc}) only works in the elastic
region. For this partial wave one needs a more dedicated method to
study, including  coupled-channel unitarity. For the poles of the
$\sigma$, the differences between the different Cases is a also bit
larger than those of other resonances such as the $\rho(770)$ and the $f_0(980)$.
This is because the $\sigma$ is far away from the real axis. For the
virtual state in the isospin-two S-wave, the poles and residues are a bit
different from Cases~A and B to Case~C. This situation is comparable
with that of $T^2_S(0)$, where in Cases~A and B $T^2_S(0)$ is 0.055 and
in Case~C it is 0.060, respectively.

\begin{figure}[t]
\includegraphics[width=0.48\textwidth,height=0.25\textheight]{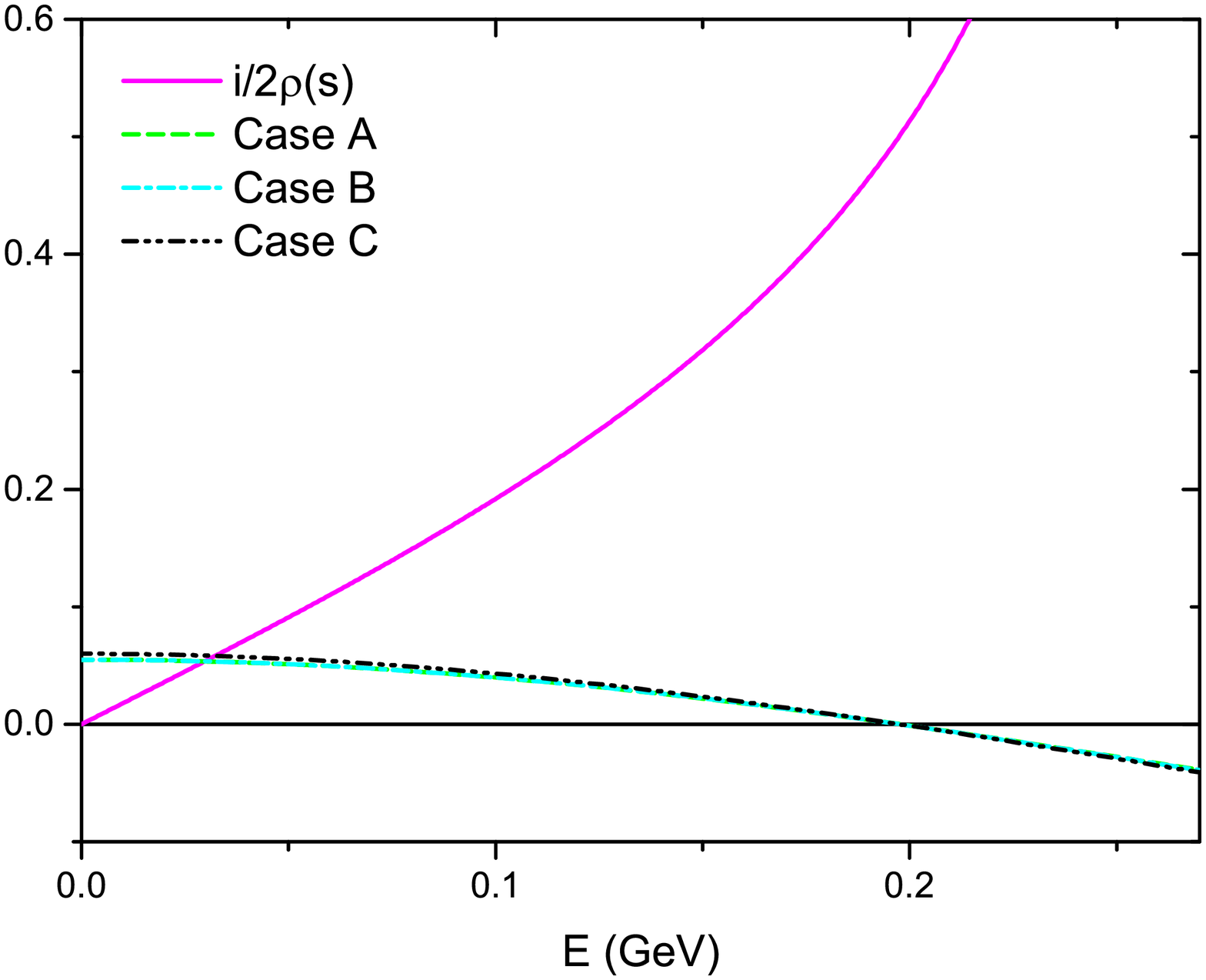}
\caption{\label{Fig:T;2S} The lines of $T^2_S(s)$ for the different Cases and
  $i/2\rho(s)$. Note that all of them are real.
The intersection point corresponds to the virtual state. }
\end{figure}

In addition, we also find that there exists a virtual
state in the isospin-two S-wave very close to $s=0$~\footnote{It has already been
discussed in \cite{Ang:2001bd}, within  a unitarized $\chi$PT method.
Here we use dispersion approach and re-confirm it, but we do
not have the extra poles caused by unitarization.}.
According to Eq.~(\ref{eq;T;II}), the virtual state a the zero of the S-matrix
below the threshold. This zero equals to the intersection point
between two lines: $T^2_S(s)$ and $i/2\rho(s)$. As shown in
Fig.~\ref{Fig:T;2S}, the line of the $T^2_S(s)$ and the line of
$i/2\rho(s)$ will always intersect with each other and the crossing
point always lies in the energy region of [0,$s_a$], where $s_a$ is the Adler zero. This is the
virtual state. Since the scattering length is negative and the Adler
zero (only one) is below threshold, one would expect that the
amplitude of $T^2_S(s)$, from $s=4 M_\pi^2$ to $s=0$, will always
cross the real axis of $s$ and arrive at the positive vertical axis.
In all events, it will intersect with that of the $i/2\rho(s)$. Thus
the existence of the virtual state is confirmed. This inference is
model-independent, only the sign of the scattering length,
\footnote{Recently, Lattice
QCD gives negative scattering length as
$-0.0412(08)(16)M_\pi^{-1}$~\cite{Beane:2011sc},
$-0.04430(25)(40)M_\pi^{-1}$~\cite{Fu:2013ffa}.
$-0.04430(2)(^{+4}_{-0})M_\pi^{-1}$\cite{Helmes:2015gla}
These values are consistent with that of $\chi$PT~\cite{Bijnens:2014lea},
the Roy equations matched to $\chi$PT~\cite{Colangelo:2000jc} and a
dispersive analysis~\cite{KPY}.}
the Adler zero, and analyticity are relevant.
For a general discussion of the virtual state arising from a bare discrete state
in the quantum mechanical scattering, we recommend readers to
read~\cite{Xiao2016,Xiao:2016mon} and references therein.
We suggest that the isospin-two S-wave amplitude could be checked in the future
measurement of $\Lambda_c^+\to\Sigma^-\pi^+\pi^+$. Its branching
ratio \cite{Ablikim:2017iqd} is large enough.

The average values of the poles and residues of all the
Cases define our central values. The deviations of the different Cases to
the central values are used to estimate the uncertainties. The results are
shown in Tab.~\ref{tab:poles}.
\begin{table}[hpb]
{\footnotesize
\begin{tabular}{|@{}c @{} | c |c |@{}c @{}|@{}c@{} |}
\hline
\rule[-0.4cm]{0cm}{0.8cm}\multirow{2}{*}{\rule[-0.8cm]{0cm}{1.6cm}State}  &  pole locations  &\multicolumn{2}{c|}{$g_{f\pi\pi}=|g_{f\pi\pi}|e^{i\phi}$~}  \\
\cline{2-4}
\rule[-0.4cm]{0cm}{0.8cm}\multirow{2}{*}{\rule[-0.8cm]{0cm}{1.6cm}}   & (MeV)   & $~|g_{f\pi\pi}|~(GeV)~$ & $~\phi~(^\circ)~$   \\
\hline\hline
$\sigma/f_0(500)$  & $437.8(52)-i270.3(5)$     & $0.47(1)$   & $-75(2)$     \rbox \\[0.5mm] \hline
$f_0(980)$         & $997.6(1)-i20.3(13)$      & $0.26(1)$   & $-82(1)$     \rbox \\[0.5mm] \hline
$f_2(1270)$        & $1273.7(179)-i89.5(280)$   & $0.53(2)$   & $-3(12)$    \rbox \\[0.5mm] \hline
$\rho(770)$        & $761.8(11) -i71.9(14) $    & $0.34(1)$   & $-12(1)$      \rbox \\[0.5mm] \hline
$2S~~v.s.$         & $30.6(15)  $    & $10.2(7)\times10^{-3}$   & $90(0)$      \rbox \\[0.5mm] \hline
\hline
\end{tabular}
\caption{\label{tab:poles}The pole locations and residues by taking
the averages between Cases~A, B and C as discussed in the text. }
}
\end{table}
These are very similar from those of previous
analyses~\cite{caprini06,DLY-MRP14,PelaezPRL,Rusetsky2011,PDG16}.
The $f_2(1270)$ has a much larger uncertainty compared to the other resonances, just as
discussed before. The residues of all resonances have roughly similar magnitude at the region
[0.25,0.55]~GeV, except for that of the virtual state in the isospin-two S-wave, which is much
weaker. But their phases are quite different. The phases
of $\rho(770)$ and $f_2(1270)$ are close to zero, while those of the $\sigma$ and $f_0(980)$
are close to  $-90^\circ$, and the virtual state one is close to $90^\circ$. This may imply
that $\rho(770)$ and $f_2(1270)$ are normal
$\bar{q}q$ states but that the $\sigma$ and $f_0(980)$  have large molecular components.

\subsection{The correlation between poles and cuts}
\begin{figure}[hpb]
\includegraphics[width=0.23\textwidth,height=0.15\textheight]{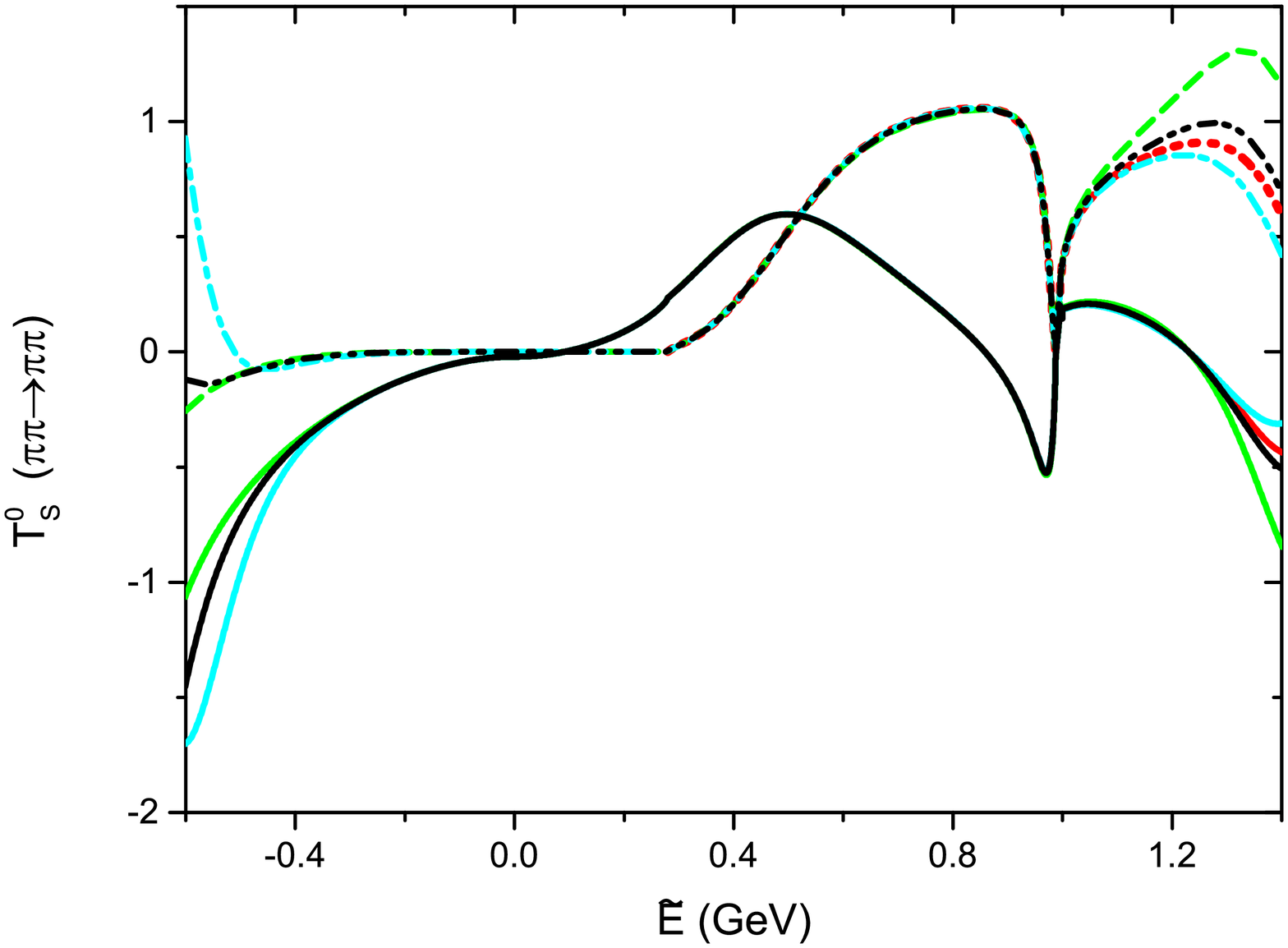}
\includegraphics[width=0.23\textwidth,height=0.15\textheight]{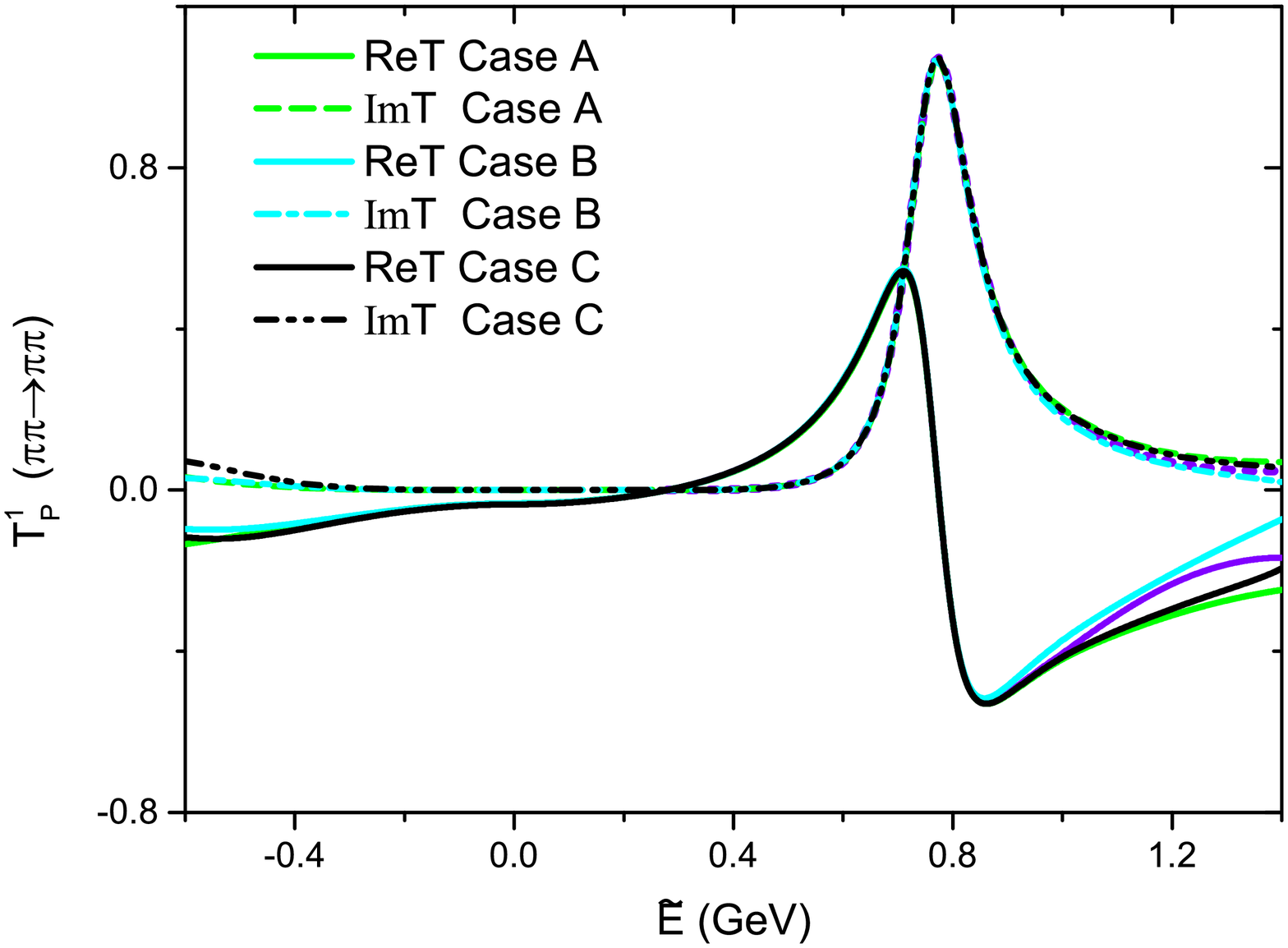}
\caption{\label{Fig:T;lhc} Comparison of different solutions of the $\pi\pi$ scattering amplitudes.
  The solid lines are the real part of the amplitudes and other lines are the imaginary part.
  The green lines are from Case A, the cyan lines are from Case B,  and the black lines are from Case C. The red
   lines are from K-Matrix \cite{DLY-MRP14} for isospin 0 S-wave, and the violet lines are from  CFDIV~\cite{KPY} for isospin 1 P-wave.
   Note that the lines of K-matrix and/or CDFIV are overlapped with our fits in the elastic region, or even a bit further in the inelastic region. }
\end{figure}

It is interesting to find the correlation between the poles and cuts.
We focus here on the isospin-zero S-wave and isospin-one P-wave, as the $f_2(1270)$ is far
away from the l.h.c and the virtual state is too close to the l.h.c..
Also, the light scalars are more difficult to  understand. All the fits of different Cases about these
two partial waves are shown in Fig.~\ref{Fig:T;lhc}.
In our approach only unitarity is used to constrain the amplitudes, but the low-energy amplitudes are
consistent with those of $\chi$PT. Only in Cases~B and for the isospin-zero S-wave, the
$T^0_S(s)$ amplitude
is inconsistent with that of $\chi$PT at $\tilde{E}<-0.4$~GeV.  This implies that unitarity has a
strong constraint on the low-energy amplitudes below $\tilde{E}=0$.
This could also be simply checked by using Eq.~(\ref{eq:Delta;T}), with $\tilde{E}=-0.1$ to
$-0.4$~GeV and $s_n=-\tilde{E}_n^2$.
Here, $\Delta T^I_J(s_n)$ is the difference between our amplitude and that of SU(2) $\chi$PT.
Typically, in Case~B, $\mathcal{R}^L_{ T^0_S}=21\%$, $\mathcal{R}^L_{ T^1_P}=16\%$,
and they are quite close to those of other Cases.
Note that the real parts of the amplitudes are more consistent with those of $\chi$PT, while the
imaginary parts have a bit larger deviation (as is expected as imaginary parts start later in the
chiral expansion).

To see the variation of the l.h.c. in the different solutions, we apply Eq.~(\ref{eq:Delta})
on ${\rm Im}T^I_J(s_n)$, with $\tilde{E}=-0.1$ to $-0.6$~GeV. Notice that at $\tilde{E}=0$ all
of the l.h.c are zero and behave as $\sqrt{-s^3}$, this partly ensures the l.h.c to be consistent with
that of  $\chi$PT in the low-energy region. We fix the average value of all Cases as the central value,
and calculate the relative deviation for each point. At last we avarage these relative deviations
to estimate the variation of the cuts. The variation of cuts and poles are defined as
\bea
\mathcal{R}_{{\rm Im} T^I_J}&=&\frac{1}{N}\sum_{n=1}^N\frac{|\Delta {\rm Im}T^I_J(s_n)|}{|{\rm Im}T^I_J(s_n)|}\;,\nonumber\\
\mathcal{R}_{pole}&=&\frac{|\Delta {\rm Re}\sqrt{s_{p}}|+|\Delta {\rm Im}\sqrt{s_{p}}|}{|\sqrt{s_{p}|}}\;,\label{eq:Delta}
\eea
with $s_p$ the pole on the second Riemann Sheet.
Finally, we collect the uncertainties  in Tab.~\ref{tab:para;C}.
And we define the correlation between poles and cuts as
\bea
C_{pole}=\frac{\mathcal{R}_{pole}}{\mathcal{R}_{{\rm Im}T^I_J}}\,.\label{eq:C}
\eea
The simple meaning of the correlation $C_{pole}$ is to answer the following question:
When the cut is changed by 100\%, how much would the pole location be changed?
\begin{table}[htbp]
\hspace{-1.5cm}
\vspace{-0.0cm}
{\footnotesize
\begin{center}
\tabcolsep=0.11cm
\begin{tabular}  {|c|c|c|c|c|c| }
\hline
                                              &                                                & Case A    &  Case B   & Case C   \rbox \\[0.5mm] \hline
\multirow{5}{*}{\rule[-1.5cm]{0cm}{3cm}l.h.c.} & $\mathcal{R}_{{\rm Im_L}T^0_S}$ & 171\%  &  126\%    &  45\%  \rbox \\[0.5mm]
\multirow{5}{*}{\rule[-1.5cm]{0cm}{3cm}}      & $C^L_{\sigma}$                              & 2.01\%    &  1.70\%   &  2.87\%  \rbox \\[0.5mm]
\multirow{5}{*}{\rule[-1.5cm]{0cm}{3cm}}      & $C^L_{f_0(980)}$                            & 0.08\%    &  0.10\%   &  0.04\%  \rbox  \\ \cline{2-5}
\multirow{5}{*}{\rule[-1.5cm]{0cm}{3cm}}      & $\mathcal{R}_{{\rm Im_L}T^1_P}$  & 41\%      &  20\%     &  61\%   \rbox \\[0.5mm]
\multirow{5}{*}{\rule[-1.5cm]{0cm}{3cm}}      & $C^L_{\rho}$                                & 0.63\%    & 1.75\%    &  1.41\%  \rbox \\[0.5mm] \hline
\multirow{5}{*}{\rule[-1.5cm]{0cm}{3cm}r.h.c.} & $\mathcal{R}_{{\rm Im_R}T^0_S}$ & 1.70\%  &  0.64\%   &  1.36\%  \rbox \\[0.5mm]
\multirow{5}{*}{\rule[-1.5cm]{0cm}{3cm}}      & $C^R_{\sigma}$                              & 387\%    &  328\%   &  189\%  \rbox \\[0.5mm]
\multirow{5}{*}{\rule[-1.5cm]{0cm}{3cm}}      & $C^R_{f_0(980)}$                            & 30\%    &  142\%   &  55\%  \rbox  \\ \cline{2-5}
\multirow{5}{*}{\rule[-1.5cm]{0cm}{3cm}}      & $\mathcal{R}_{{\rm Im_R}T^1_P}$  & 6.6\%     &  10.0\%   &  4.4\%   \rbox \\[0.5mm]
\multirow{5}{*}{\rule[-1.5cm]{0cm}{3cm}}      & $C^R_{\rho}$                                & 17.6\%    &  6.4\%  &  28.2\%  \rbox \\[0.5mm] \hline
\hline
\hline
\end{tabular}
\caption{\label{tab:para;C} The correlation between l.h.c. and poles (represented by superscript \lq L'), and  between r.h.c. and poles (represented by superscript \lq R').  }
\end{center}
}
\end{table}

To test the correlation between poles and the r.h.c., we simply set $\varphi_{test}(s)
= 1.04\varphi(s)$, and check the variation of poles and cuts, respectively.
The relative uncertainty of the r.h.c. is also estimated by  Eq.~(\ref{eq:Delta}),
with $s_n=0.1-0.9$~GeV$^2$ for the isospin-zero S-wave and $s_n=0.1-0.8$~GeV$^2$ for the
isospin-one P-wave. The relative uncertainty of the poles and the correlation are calculated
in the same way as that of the l.h.c, see Eqs.~(\ref{eq:Delta},\ref{eq:C}).

From  Tab.~\ref{tab:para;C}, we find that $C_\sigma^R$ is roughly two orders larger than that
of $C_\sigma^L$, though $\sigma$ is rather close to the l.h.c..
Comparing to Ref.~\cite{caprini06}, which has roughly 15\% contribution from l.h.c, we have a rather
smaller contribution from the l.h.c, caused by the constraint of unitarity on the l.h.c.\ .
Also, $C_{f_0(980)}^R$ is roughly three orders larger than that of $C_{f_0(980)}^L$, and
$C_{\rho(770)}^R$ is roughly one order larger than that of $C_{\rho(770)}^L$.
These indicate that the correlation between the unitarity cut and the poles is much larger than
that of the l.h.c. and poles. Note that in our case the l.h.c. is not arbitrary but correlated with
the r.h.c., constrained by unitarity and analyticity, see Eq.~(\ref{eq:unit;lhc}).
For each Case, $C_\sigma^{L}$ is larger than $C_{f_0(980)}^{L}$. This is not surprising as the $\sigma$ is
much  closer to the l.h.c.\ . Also, $C_\sigma^{R}$ is larger than $C_{f_0(980)}^{R}$. The reason is
that the $\sigma$ is farther away from the real axis, the uncertainty of the pole is larger as
the amplitude is continued from the physical region  deeper into the complex-$s$ plane.
It is interesting to see that in average $C_\sigma^{L}$ is roughly two times larger than $C_{\rho(770)}^L$.
And for the distance between these poles and l.h.c (simply set $s=0$),  $|s_\sigma|$ is one half of that
of $|s_\rho|$, this tells us that the correlation between poles and l.h.c is inversely proportional to
their distance. In contrast, $C_\sigma^{R}$ is roughly one order larger than $C_{\rho(770)}^R$.
For the distance between these poles and r.h.c. (simply set $s={\rm Re}~s_{pole}$),
$|{\rm Im}~s_\sigma|$ is two times larger than $|{\rm Im}~s_\rho|$, this tells us that the
correlation between poles and r.h.c. is proportional to their distance.
These conclusions are still kept when comparing the $\sigma$ and the $f_0(980)$.

\section{Summary}\label{sec:summary}
We proposed a dispersive method to calculate the two-body scattering amplitude. It is
based on the Omn\`es function of the phase, including that of the left hand cut
and the right hand cut. The input of the r.h.c. is given by three kinds of parametrizations,
and the l.h.c is solved by Eq.~(\ref{eq:unit;lhc}), with unitarity and analyticity respected.
The pion-pion $IJ=00,02,11,20$ waves are fitted within our method and the
poles and locations are extracted. They are stable except for that of the $f_2(1270)$, which lies in
the inelastic region. The r.h.c. has much larger contribution to the poles comparing to that
of the l.h.c.. This method could be useful for the studies of strong interactions in
two-body scattering, and the $\pi\pi$ scattering amplitudes obtained here could be used for the
future studies when one has $\pi\pi$ final state
interactions, see e.g.~\cite{AMP-FSI,Dai:2012pb,Gonzalez-Solis:2018xnw,Ropertz:2018stk,Cheng:2019hpq},
and/or to multi-pions, see e.g.~ \cite{Guo:2015zqa,Dumm:2009va}.

\section*{Acknowledgements}
We are grateful to Zhi-Yong Zhou for helpful discussions and for supplying us the files of the
$SU(2)$ 2-loop $\chi$PT amplitudes. This work is supported by National Natural Science Foundation of
China (NSFC) with Grant Nos.11805059, 11805012, 11805037, and Fundamental Research Funds for the Central Universities.
TL also thanks support from the Joint Large Scale Scientific Facility Funds of the NSFC and Chinese Academy of Sciences (CAS)  under Contract No. U1832121,
and from Shanghai Pujiang Program under Grant No.18PJ1401000,  Open Research Program of Large Research Infrastructures (2017), CAS.
UGM acknowledges support from the DFG (SFB/TR 110, ``Symmetries and the Emergence of Structure in QCD''),
from the Chinese Academy of Sciences (CAS) President's International Fellowship Initiative (PIFI) (Grant No. 2018DM0034)
and from VolkswagenStiftung (Grant No. 93562).

\appendix
\setcounter{equation}{0}
\setcounter{table}{0}
\renewcommand{\theequation}{\Alph{section}.\arabic{equation}}
\renewcommand{\thetable}{\Alph{section}.\arabic{table}}

\section{Analytical amplitudes of partial waves within $\chi$PT}\label{app:ChPT}
The analytical 1-loop amplitudes of these partial waves within of $SU(3)$ $\chi$PT are
recalculated. For reader's convenience, they are given below.
We have the IJ=00 waves up to $\mathcal{O}(p^4)$:
\begin{widetext}
\bea
T^{(2)}_{0S}[s]&=&\frac{2 s- M_{\pi}^2}{32 \pi  f_{\pi}^2}\;,\\[3mm]
t_{0S,1}[s]&=&\frac{1}{3}(3 s-4 M_{\pi}^2) A[M_K]+\frac{2}{3} (3 s-4 M_{\pi}^2) A[M_{\pi}]
+\frac{1}{6} M_{\pi}^4 B[s,M_{\eta }]+\frac{3}{8} s^2 B[s,M_K]\no\\
&&+\frac{1}{2} (2 s-M_{\pi}^2)^2B[s,M_{\pi}]+\frac{128\pi^2}{3}(-40M_{\pi}^2 s+44 M_{\pi}^4+11 s^2)L_1\no\\
&&+\frac{128\pi^2}{3} (-20 M_{\pi}^2 s+28 M_{\pi}^4+7 s^2)L_2+\frac{64\pi^2}{3}(-40M_{\pi}^2 s+44 M_{\pi}^4+11 s^2) L_3\no\\
&&+256 \pi ^2 M_{\pi}^2 (s-3M_{\pi}^2)(2L_4+L_5)+1280\pi^2 M_{\pi}^4 (2L_6+L_8)\no\\
&&-\frac{1}{9}(9 s M_K^2-12 M_{\pi}^2 M_K^2+8 M_{\pi}^2 s-16 M_{\pi}^4+2 s^2)\;,\\[3mm]
t_{0S,2}[s]&=&-\left\{\frac{5}{12} H_2[4M_{\pi}^2-s,M_K]+2(\frac{1}{6} s-\frac{1}{3}M_K^2-\frac{1}{3}M_{\pi}^2)
H_1[4M_{\pi}^2-s,M_K]+\frac{10}{3} H_2[4M_{\pi}^2-s,M_{\pi}]\right.\no\\
&&+2(-\frac{2}{3} s M_K^2+\frac{4}{3} M_K^2 M_{\pi}^2)H_0[4M_{\pi}^2-s,M_K]+2(\frac{1}{3}s -\frac{16}{3} M_{\pi}^2)
H_1[4M_{\pi}^2-s,M_{\pi}]\no\\
&&\left.+2(-\frac{4}{3} s M_{\pi}^2+\frac{37}{6} M_{\pi}^4)H_0[4M_{\pi}^2-s,M_{\pi}]+\frac{1}{9}M_{\pi}^4H_0[4M_{\pi}^2-s,M_{\eta }]\right\}\;,\\[3mm]
T^{(4)}_{0S}[s]&=&\frac{1}{32 \pi  f_{\pi }^2}\frac{1}{16 \pi ^2 f_\pi^2}\left(t_{0S,1}[s]+\frac{t_{0S,2}[s]}{s-4M_{\pi}^2}\right)\;.\label{eq:ChPT;0S}
\eea
Here the superscript of $T$ in the bracket means the chiral order, and subscripts represent for Isospin and spin, respectively. Note that for reader's convenience we also give the analytical forms of the imaginary part (r.h.c.) of the amplitudes.
The I=2 S wave is
\bea
T^{(2)}_{2S}[s]&=&-\frac{s-2 M_{\pi}^2}{32 \pi  f_{\pi }^2}\;,\\[3mm]
t_{2S,1}[s]&=&(\frac{2}{3} M_{\pi}^2-\frac{1}{2}s) A[M_K]+(\frac{4}{3} M_{\pi}^2-s) A[M_{\pi}]+\frac{1}{2} B[s,M_{\pi}] (s-2 M_{\pi}^2)^2\no\\
&&+\frac{1}{18}(9 s M_K^2-12 M_{\pi}^2 M_K^2-16 M_{\pi}^4+2 s^2+8 M_{\pi}^2 s)+\frac{256\pi^2}{3} (-2 M_{\pi}^2 s+4 M_{\pi}^4+s^2)L_1\no\\
&&+\frac{256\pi ^2}{3} (-7 M_{\pi}^2 s+8 M_{\pi}^4+2 s^2) L_2-128 \pi^2 M_{\pi}^2 s (2L_4+L_5)\no\\
&&+\frac{128\pi ^2}{3} (-2 M_{\pi}^2 s+4 M_{\pi}^4+ s^2)L_3+512 \pi ^2 M_{\pi}^4(2L_6+L_8)\;,\\[3mm]
t_{2S,2}[s]&=&-2\left\{\frac{1}{12}H_2[4M_{\pi}^2-s,M_K]+(\frac{M_K^2}{6}+\frac{M_{\pi}^2}{6}-\frac{s}{12})H_1[4M_{\pi}^2-s,M_K]\right.\no\\
&&+(\frac{s M_K^2}{3}-\frac{2}{3} M_{\pi}^2 M_K^2)H_0[4M_{\pi}^2-s,M_K]+\frac{2}{3}H_2[4M_{\pi}^2-s,M_{\pi}]
+(-\frac{M_{\pi}^2}{3}-\frac{s}{6})H_1[4M_{\pi}^2-s,M_{\pi}]\no\\
&&\left.+(\frac{2M_{\pi}^2 s}{3}-\frac{5 M_{\pi}^4}{6})H_0[4M_{\pi}^2-s,M_{\pi}]+\frac{M_{\pi}^4}{18}H_0[4M_{\pi}^2-s,M_{\eta }]\right\}\;,\\[3mm]
T^{(4)}_{2S}[s]&=&\frac{1}{32\pi  f_{\pi }^2}\frac{1}{16 \pi ^2 f_{\pi }^2}\left(t_{2S,1}[s]+\frac{t_{2S,2}[s]}{s-4M_{\pi}^2}\right)\;. \label{eq:ChPT;2S}
\eea
The I=1 P wave is
\bea
T^{(2)}_{1P}[s]&=&\frac{s-4 M_{\pi}^2}{96 \pi  f_{\pi }^2}\;,\\[3mm]
t_{1P,1}[s]&=&\frac{1}{36} (s-4 M_{\pi}^2) \left\{ 3\left[2 A[M_K]+4 A[M_{\pi}]-2
M_K^2+4 M_{\pi}^2 \left(128 \pi ^2 (2 L_4+L_5)-1\right)\right.\right.\no\\
&&\left.\left.-256 \pi ^2 s (2 L_1-L_2+L_3)+s\right]+B[s,M_K](s-4M_K^2)+2B[s,M_{\pi}] (s-4 M_{\pi}^2)\right\}\;,\\[3mm]
t_{1P,2}[s]&=&-\left\{\frac{2}{3}H_3[4M_{\pi}^2-s,M_K]+\left(\frac{1}{6}(2s-4M_K^2-4 M_{\pi}^2)+\frac{1}{3}(s-4M_{\pi}^2)\right) H_2[4M_{\pi}^2-s,M_K]\right.\no\\
&&+\left(\frac{1}{6} (-8 s M_K^2+16 M_K^2 M_{\pi}^2)+\frac{1}{12}(s-4M_{\pi}^2)(2s-4M_K^2-4 M_{\pi}^2)\right)H_1[4M_{\pi}^2-s,M_K]\no\\
&&+\frac{1}{12}(s-4M_{\pi}^2)(-8 s M_K^2+16 M_K^2 M_{\pi}^2)H_0[4M_{\pi}^2-s,M_K]+\frac{4}{3}H_3[4M_{\pi}^2-s,M_{\pi}]\no\\
&&+\left(\frac{2}{3}(s+2 M_{\pi}^2)+\frac{2}{3}(s-4M_{\pi}^2)\right) H_2[4M_{\pi}^2-s,M_{\pi}]+\frac{2M_{\pi}^4}{9}H_1[4M_{\pi}^2-s,M_{\eta }]\no\\
&&+\left(-\frac{2}{3}(4 s M_{\pi}^2+M_{\pi}^4)+\frac{1}{3}(s+2 M_{\pi}^2)(s-4M_{\pi}^2)\right) H_1[4M_{\pi}^2-s,M_{\pi}]\no\\
&&\left.-\frac{1}{3}(4 s M_{\pi}^2+M_{\pi}^4)(s-4M_{\pi}^2)H_0[4M_{\pi}^2-s,M_{\pi}]+\frac{M_{\pi}^4}{9}(s-4m_{\pi
}^2)H_0[4M_{\pi}^2-s,M_{\eta }]\right\}\;,\\[3mm]
T^{(4)}_{1P}[s]&=&\frac{1}{32\pi f_\pi^2 }\frac{1}{16 \pi ^2 f_\pi^2}\left(t_{1P,1}[s]+\frac{t_{1P,2}[s]}{(s-4M_{\pi}^2)^2}\right)\;.\label{eq:ChPT;1P}
\eea
And the I=0 D wave is
\bea
T^{(2)}_{0D}[s]&=&0\;,\\[3mm]
t_{0D,1}[s]&=&\frac{1}{90}(s-4 M_{\pi}^2)^2 \left(384 \pi ^2 (2 L_1+4 L_2+L_3)+1\right)\;,\\[3mm]
t_{0D,2}[s]&=&-2\left\{\frac{5}{4}H_4[4M_{\pi}^2-s,M_K]+\left((-2M_K^2-2M_{\pi}^2+s)+\frac{5}{4}(s-4M_{\pi}^2)\right)H_3[4M_{\pi}^2-s,M_K]\right.\no\\
&&+\left((8 M_{\pi}^2 M_K^2-4 s M_K^2)+(s-4M_{\pi}^2)(-2M_K^2-2M_{\pi}^2+s)+\frac{5}{24}(s-4M_{\pi}^2)^2\right)H_2[4M_{\pi}^2-s,M_K]\no\\
&&+\left((s-4M_{\pi}^2)(8 M_{\pi}^2 M_K^2-4 s M_K^2)+(s-4M_{\pi}^2)^2(-\frac{M_K^2}{3}-\frac{M_{\pi}^2}{3}+\frac{s}{6})\right)H_1[4M_{\pi}^2-s,M_K]\no\\
&&+(s-4M_{\pi}^2)^2(\frac{4}{3} M_{\pi}^2 M_K^2-\frac{2 s M_K^2}{3})H_0[4M_{\pi}^2-s,M_K]+10 H_4[4M_{\pi}^2-s,M_{\pi}]\no\\
&&+\left((2s-32M_{\pi}^2)+10(s-4M_{\pi}^2)\right)H_3[4M_{\pi}^2-s,M_{\pi}]\no\\
&&+\left((37 M_{\pi}^4-8M_{\pi}^2 s)+(s-4M_{\pi}^2)(2s-32M_{\pi}^2)+\frac{5}{3}(s-4M_{\pi}^2)^2\right)H_2[4M_{\pi}^2-s,M_{\pi}]\no\\
&&+\left((s-4M_{\pi}^2)(37 M_{\pi}^4-8M_{\pi}^2 s)+(s-4M_{\pi}^2)^2(\frac{s}{3}-\frac{16 M_{\pi}^2}{3})\right)H_1[4M_{\pi}^2-s,M_{\pi}]\no\\
&&+(s-4M_{\pi}^2)^2(\frac{37 M_{\pi}^4}{6}-\frac{4 M_{\pi}^2 s}{3})H_0[4M_{\pi}^2-s,M_{\pi}]
+\frac{M_{\pi}^4}{3}H_2[4M_{\pi}^2-s,M_{\eta }]\no\\
&&\left.+(s-4M_{\pi}^2)\frac{M_{\pi}^4}{3}H_1[4M_{\pi}^2-s,M_{\eta }]+\frac{M_{\pi}^4}{18}(s-4M_{\pi}^2)^2
H_0[4M_{\pi}^2-s,M_{\eta }]\right\}\;,\\[3mm]
T^{(4)}_{0D}[s]&=&\frac{1}{32\pi  f_{\pi }^2}\frac{1}{16 \pi ^2 f_{\pi }^2}\left(t_{0D,1}[s]+\frac{t_{0D,2}[s]}{(s-4M_{\pi}^2)^3}\right)\;.\label{eq:ChPT;0D}
\eea
It should be noted that in all these partial waves, $2L_4+L_5$ and  $2L_6+L_8$ appear together \cite{DLY11}.
The $A$, $B$, $H$ functions are given as below
\bea
A[m]&=&m^2\left(1-\ln\frac{m^2}{\mu ^2}\right)\;,\\
B[s,m]&=&2-\ln\frac{m^2}{\mu ^2}-\rho (s,m)\ln\left(\frac{\rho (s,m)+1}{\rho (s,m)-1}\right)\;,\\
H_{0}[t,m]&=&-t \ln\frac{m^2}{\mu ^2}+m^2 \ln^2\left(\frac{\rho (t,m)+1}{\rho (t,m)-1}\right)- t
\rho (t,m)\ln\left(\frac{\rho (t,m)+1}{\rho (t,m)-1}\right)+3 t \;,\\
H_{1}[t,m]&=&\frac{1}{4} t^2 \left(5-2 \ln\frac{m^2}{\mu ^2}\right)-m^2 t+m^4 \ln^2\left(\frac{\rho (t,m)+1}{\rho
(t,m)-1}\right)-\frac{1}{2} t (t-2 m^2) \rho(t,m)\ln\left(\frac{\rho (t,m)+1}{\rho (t,m)-1}\right)\;,\\
H_{2}[t,m]&=&\frac{1}{9} t^3 \left(7-3 \ln\frac{m^2}{\mu ^2}\right)-\frac{1}{6} m^2 t^2-\frac{1}{3} t \rho(t,m) (-t
m^2 -6 m^4+t^2) \ln\left(\frac{\rho(t,m)+1}{\rho(t,m)-1}\right)\no\\
&&-2 m^4 t+2 m^6 \ln^2\left(\frac{\rho(t,m)+1}{\rho(t,m)-1}\right)\;,\\
H_{3}[t,m]&=&-\frac{1}{12} t \rho(t,m) (-2 t^2 m^2-10 t m^4-60 m^6+3 t^3) \ln\left(\frac{\rho (t,m)+1}{\rho (t,m)-1}\right)\nonumber\\
&&+\frac{1}{16} t^4 \left(9-4 \ln\frac{m^2}{\mu ^2}\right)-\frac{1}{18} t^3 m^2-\frac{5}{12} t^2
m^4-5 t m^6+5 m^8 \ln^2\left(\frac{\rho(t,m)+1}{\rho(t,m)-1}\right)\;,\\
H_{4}[t,m]&=&\frac{1}{25} t^5 \left(11-5 \ln\frac{m^2}{\mu ^2}\right)-\frac{1}{40} t^4 m^2-\frac{7}{45} t^3 m^4-\frac{7}{6}
t^2 m^6-14 t m^8+14 m^{10} \ln^2\left(\frac{\rho(t,m)+1}{\rho(t,m)-1}\right)\nonumber\\
&&-\frac{1}{30}t \rho(t,m) (-3 t^3 m^2-14 t^2 m^4-70 t m^6-420 m^8+6 t^4) \ln\left(\frac{\rho(t,m)+1}{\rho(t,m)-1}\right)\;,
\eea
with $\rho(t,m)=\sqrt{1-4 m^2/t}$. Notice that
our amplitudes are calculated in the formalism of $\overline{MS}$, while that of \cite{Gasser1984,Pelaez02,Guo2011} is done in $\overline{MS}-1$. The relation between our LECs ($L_i$) and that of the latter one ($\tilde{L}_i$) is $\tilde{L}_i=L_i+\frac{\Gamma_i}{32\pi^2}$.
\end{widetext}


\end{document}